\documentstyle[twocolumn,aps,prl,epsf,floats]{revtex}
\def\half{\mbox{$1\over 2$}}

\def\beq{\begin{equation}}
\def\eeq{\end{equation}}

\def\bq{\begin{equation}}
\def\eq{\end{equation}}
\def\beqa{\begin{eqnarray}}
\def\eeqa{\end{eqnarray}}

\def\LP{\left(}
\def\RP{\vphantom{\half} \right)}

 \gdef\aver#1{\left\langle #1 \right\rangle}
 \gdef\s#1{\! #1 \!}
 
 \gdef\Eq#1{Eq.~(\ref{#1})}
 
 \gdef\vec#1{{\bf #1}} 

\begin{document}
\preprint{NSF-ITP-98-032}

\title{Density Functional Theory --- an introduction}
\author{Nathan Argaman$^{1,2}$ and Guy Makov$^2$}
\address{$^1$ Institute for Theoretical Physics,
University of California, Santa Barbara, CA 93106, USA }
\address{$^2$ Physics Department, NRCN, P.O.\ Box 9001, 
Beer Sheva 84190, Israel}


\address{\parbox{14.5cm}{\rm\small
\medskip\bigskip
Density Functional Theory (DFT) is one of the most widely used methods
for ``ab initio'' calculations of the structure of atoms, molecules,
crystals, surfaces, and their interactions.  Unfortunately, the 
customary introduction to DFT is often considered too lengthy to be 
included in various curricula.  An alternative introduction to DFT is 
presented here, drawing on ideas which are well--known from 
thermodynamics, especially the idea of switching between different 
independent variables.  The central theme of DFT, 
i.e.\ the notion that it is possible and beneficial to replace the 
dependence on the external potential $v(\vec{r})$ by a dependence on 
the density distribution $n(\vec{r})$, is presented as a 
straightforward generalization of the familiar Legendre 
transform from the chemical potential $\mu$ to the number of 
particles $N$.  This approach is used here to introduce the 
Hohenberg--Kohn energy functional and to obtain the 
corresponding theorems, using classical nonuniform fluids as 
simple examples.  The energy functional 
for electronic systems is considered next, and the Kohn--Sham 
equations are derived.  The exchange--correlation part of this 
functional is discussed, including both the local density 
approximation to it, and its formally exact expression in terms of 
the exchange--correlation hole.  A very brief survey of various 
applications and extensions is included.
}}
\maketitle

\section{Introduction}

\begin{figure}[t]
\epsfxsize=\hsize 
\centerline{\epsffile{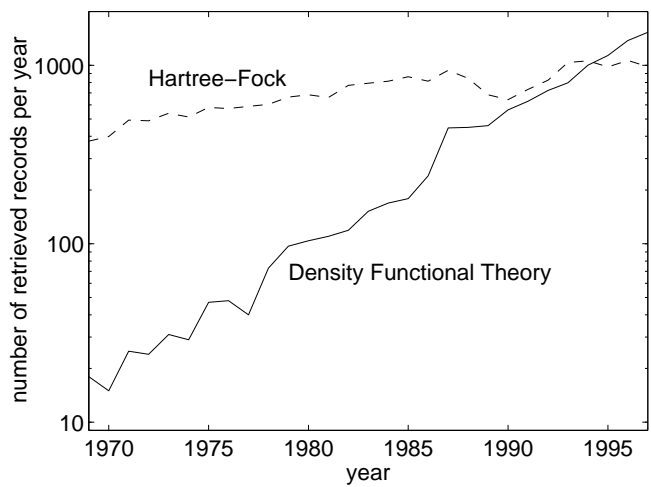}}
\vspace{.3cm}
\caption{ One indicator of the increasing use of DFT is the 
number of records retrieved from the INSPEC databases by 
searching for the keywords ``density'', ``functional'' and 
``theory''. This is compared here with a similar search for 
keywords ``Hartree'' and ``Fock'', which parallels the overall 
growth of the INSPEC databases (for any given year, approximately 
0.3\% of the records have the Hartree--Fock keywords). }
\label{stats}
\end{figure}

The predominant theoretical picture of solid--state and/or molecular
systems involves the inhomogeneous electron gas: a set of interacting
point electrons moving quantum--mechanically in the potential field 
of a set of
atomic nuclei, which are considered to be static (the Born--Oppenheimer
approximation).  Solution of such models generally requires the use of
approximation schemes, of which the most basic --- the independent
electron approximation, the Hartree theory and Hartree--Fock theory ---
are routinely taught to undergraduates in Physics and Chemistry
courses.  However, there is another approach  --- 
Density Functional Theory (DFT) --- which over the last thirty
years or so has become increasingly the method of choice for the
solution of such problems (see Fig.~\ref{stats}.).  
This method has the double advantage of being able to treat
many problems to a sufficiently high accuracy, as well as being 
computationally simple (simpler even than the Hartree scheme).  
Despite these advantages it is absent from most undergraduate and 
many graduate curricula with which we are familiar.

We believe that this omission stems in part from the tendency of the 
existing books and review papers on DFT, e.g.\
Refs.~\cite{ParrYang,Dreizler,Inhom}, to follow the
historical path of development of the theory.  Although appropriate 
for a thorough treatment, this approach unnecessarily
prolongs the introduction and grapples with problems which are not
directly relevant to the practitioner.  It is our purpose here to give
a brief and self--contained introduction to density functional theory,
assuming only a first course in quantum mechanics and in
thermostatistics.  We break with the traditional approach by relying
on the analogy with thermodynamics \cite{Legendre}.  In this
formulation, the use of the density distribution as a free variable 
arises in a natural manner, as do more advanced concepts which are 
central to recent developments in the theory \cite{KBP}, e.g.\ 
the exchange--correlation hole and generalized compressibilities.  
The discussion is sufficiently detailed to provide a useful
overview for the beginning practitioner, and the relatively novel 
point of view may also prove illuminating for those experienced 
researchers who are not familiar with it.  We hope that the 
availability of such an introduction will encourage teachers 
to include a one or two hour class on DFT in courses on
quantum mechanics, atomic and molecular physics, condensed matter 
physics, and materials science.

The general theoretical framework of DFT, involving the 
Hohenberg--Kohn free energy $F_{\text{HK}}[n(\vec{r})]$, is 
presented in Sec.~II, which for simplicity focuses on classical 
systems.  The generalization to the quantum--mechanical 
electron gas is given in Sec. III, together with the discussion of 
the Kohn--Sham equations and of the local density approximation, 
which is the simplest practical approximation for the 
exchange--correlation energy.  Various issues relating to the  
accuracy of this approach are discussed in Sec.~IV, followed by 
a summary in Sec.~V.

\section{General theory}

In this section, a unified treatment of thermodynamics and density
functional theory is presented.  For simplicity, the case of a
{\it classical} interacting system of point particles will be 
discussed first.  Although classical DFT has its own applications, 
e.g.\ liquids \cite{liquids}, the reader is advised to keep in mind 
electronic systems, which will be the subject of the next section.  
Thus, Eqs.~(\ref{Hamil0}) through (\ref{Euler}), to be 
derived here using classical notation, are equally applicable to 
quantum--mechanical systems, where Hilbert space with its position 
and momentum operators replaces the classical phase space and its 
scalar coordinates.

\subsection{Thermodynamics: a reminder}

We begin by rederiving the equations of thermodynamics from
statistical mechanics \cite{Callen}.
Consider a classical system of $M$ interacting particles in 
a container of volume $V$.  The many--body Hamiltonian is:
\begin{equation}  \label{Hamil0}
{\cal H}_{\text{MB}} \; = \; {\cal T} + {\cal U} \; ,
\end{equation}
where ${\cal T} = \sum_{i=1}^M \vec{p}_i^2/2m$ is the kinetic energy, 
and ${\cal U} = \sum_{i < j} u(|\vec{r}_i-\vec{r}_j|)$ is the 
interaction energy, assuming a simple pair potential $u(r)$.  
Here $\vec{r}_i$ and $\vec{p}_i$ are the positions and momenta 
of the particles, and $m$ is their mass.  We consider the
grand--canonical ensemble, where the system is in contact with a
heat reservoir of temperature $T$ and a particle reservoir with
chemical potential $\mu$.
It is well--known from statistical physics that the 
grand potential, which is the free energy in this case, 
is given by:
\begin{equation} \label{GP_def}
\Omega(\mu,T,V) \; = \; - T \, \log \, \Xi \; ,
\end{equation}
where $\Xi$ is the grand partition function,
\begin{equation} \label{Xi_def}
\Xi(\mu,T,V) = \sum_{M=0}^\infty {1 \over M!} {\rm Tr} \, \exp 
     \left( - { {\cal H}_{\text{MB}} \! - \! \mu M \over T} \right) \; ,
\end{equation}
the temperature is in energy units (i.e.\ $k_B=1$), and the
classical trace, ${\rm Tr}$, represents the $6M$--dimensional
phase--space integral (the division by $M!$ compensates for
double counting of many--body states of indistinguishable particles).

It follows directly from these definitions that the expectation
value of the number of particles in the system is given by a
derivative of the grand potential, $N = \aver{M} =
-(\partial \Omega / \partial \mu)$.
The convexity of the thermodynamic potential \cite{convex} implies
that $N$ is a monotonically increasing function of $\mu$.
Other partial derivatives of $\Omega$ give the values of 
additional physical quantities, such as the entropy, 
$S=-(\partial\Omega / \partial T)$ and the pressure $P = 
-(\partial\Omega / \partial V)$.  This may be summarized by writing
${\rm d}\Omega = -N \, {\rm d}\mu - S \, {\rm d}T - P \, {\rm d}V$.

\begin{figure}[t]
\epsfxsize=\hsize 
\centerline{\epsffile{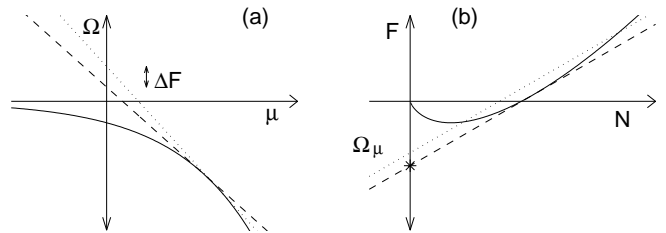}}
\vspace{.3cm}
\caption{ (a) The Legendre transform which gives $F(N)$ 
corresponds to describing 
the curve $\Omega(\mu)$ by the properties of its tangents: minus 
their slopes $N=-(\partial \Omega/\partial \mu)$ and their 
intercepts with the energy axis, $F = \Omega+\mu N$.  The fact that 
the derivative $\Delta F/\Delta N$ is equal to $\mu$ follows from 
asking the question: if two neighboring lines intercept the energy 
axis at a distance $\Delta F$ from each other, and have slopes which 
differ by $\Delta N$, how far from the axis will they cross each other?
(b) The Legendre transform back from $F(N)$ to $\Omega(\mu)$ has a 
similar interpretation.  The minimization suggested in \Eq{FEF} 
corresponds to studying a family of lines with a fixed slope $\mu$, 
which pass through points $(N,F)$ on the free energy curve.  
Their intercepts, $\Omega_\mu = F-\mu N$, have a minimum (marked by 
an asterisk) for that line which is tangent to the curve. }
\label{geometric}
\end{figure}
A basic lesson of thermodynamics is that in different contexts it is
advantageous to use different ensembles.  For example, in studying
systems where the number of particles rather than the chemical
potential is fixed, it is preferable to use the Helmholtz free energy
\cite{pointb}, which is obtained from the grand potential $\Omega$ by
a Legendre transform: $F(N,T,V) = \Omega\LP \mu(N),T,V \RP + \mu(N) N$.
Here $\mu(N)$ is no longer an independent variable, but a function 
of $N$ obtained by inverting the relationship
$N = N(\mu,V,T) = -(\partial\Omega/\partial\mu)$.
The derivative of $F$ with respect to the ``new'' free variable
$N$ is equal to the ``old'' free variable $\mu$.  The derivatives
with respect to the other variables are unchanged (but are taken
at constant $N$ rather than at constant $\mu$).  We thus write
${\rm d}F = \mu \, {\rm d}N - S \, {\rm d}T - P \, {\rm d}V$.

For the purpose of comparison with DFT, it is useful to
make a variation on the inverse Legendre transform which expresses 
$\Omega$ in terms of $F$, and to define the following ``grand 
potential function'', which depends explicitly on 
{\it both} $\mu$ and $N$:
\begin{equation}  \label{FEF}
\Omega_{\mu}(N,T,V) \; \equiv \; F(N,T,V) - \mu N \; .
\end{equation}
This function gives the original grand potential of \Eq{GP_def} when
{\it minimized} with respect to $N$, i.e.\ when the derivative
$(\partial F/\partial N) - \mu$ vanishes, which is equivalent 
to the condition $N \! = \! N(\mu,T,V)$ conventionally used   
in the inverse Legendre transform.  For other values of $N$, the 
function $\Omega_{\mu}(N,T,V)$ describes a ``cost'' in free energy of 
having a configuration with the ``wrong'' number of electrons.
For a geometric interpretation of Legendre transforms, including
the minimization procedure of \Eq{FEF}, see Fig.~\ref{geometric}.

\subsection{Nonuniform systems and the Hohenberg--Kohn theorem}

The discussion above can be generalized in a quite straightforward
manner to the treatment of particles in an external potential
$v(\vec{r})$.  The many--body Hamiltonian is now
\begin{equation}  \label{Hamil}
{\cal H}_{\text{MB}} \; = \; {\cal T} + {\cal V} + {\cal U} \; ,
\end{equation}
where the potential energy, ${\cal V} = \sum_{i=1}^M v(\vec{r}_i)$,
has been added.  The grand potential and the partition function are
defined as before, Eqs.~(\ref{GP_def}) and (\ref{Xi_def}), but they
now depend on the potential function $v(\vec{r})$ rather than on the 
scalar volume $V$.  In this sense,
$\Omega = \Omega\LP\mu,T,[v(\vec{r})]\RP$ is now a functional \cite{fnal}
of $v(\vec{r})$ as well as a function of $\mu$ and $T$ --- the square
brackets denote functional variables {\bf (}it is also implicitly a 
functional of the pair potential $u(r)${\bf )}.  As is well known, the 
potential $v(\vec{r})$, is an energy which is measured from an arbitrary 
origin, i.e.\ shifting the potential by a constant does not 
affect the physics of the system.  It is convenient here to set
this origin at the chemical potential, i.e.\ to take $\mu=0$.
Equivalently, one may define the new functional variable as
$v(\vec{r})-\mu$, as $\Omega$ depends only on this difference, and 
not on $v$ and $\mu$ separately \cite{indep}.

The functional derivative of $\Omega$ with respect to the new
variable gives the {\it density distribution} of the
particles, $n(\vec{r}) = \aver{\rho(\vec{r})} =
\delta \Omega / \delta v(\vec{r})$, where
$\rho(\vec{r}) = \sum_{i=1}^M \delta(\vec{r}-\vec{r}_i)$ is the
unaveraged density.  Using a (functional)
Legendre transform as above, we can define a new free energy
which depends on $n(\vec{r})$ rather than on $v(\vec{r})$, and is
called the Hohenberg--Kohn free energy:
\begin{equation} \label{F[n]_def}
F_{\text{HK}}[n(\vec{r})]  \; = \;  \Omega[v(\vec{r})] \, - 
    \int d\vec{r} \; n(\vec{r}) \, v(\vec{r})  \; ,
\end{equation}
where the explicit temperature variable has been omitted, and 
$v(\vec{r})$ on the right hand side is chosen to correspond to the 
given $n(\vec{r})$ (that such a choice is possible follows from 
the ``generalized convexity'' of the free energy \cite{convex,Levy}).  
The partial and functional derivatives of
$F_{\text{HK}}[n(\vec{r})]$ are given by the usual rules for
Legendre transforms: ${\mathrm d}F_{\text{HK}} = - S{\mathrm d}T 
- \int {\mathrm d}\vec{r} \> v(\vec{r}) \delta n(\vec{r})$.

The direct generalization of the free energy function of \Eq{FEF}
is the free energy functional:
\begin{equation} \label{FEFl}
\Omega _v[n(\vec{r})]  \; \equiv \;  F_{\text{HK}}[n(\vec{r})] 
\, + \int d\vec{r} \; n(\vec{r}) \, v(\vec{r})  \; ,
\end{equation}
with $v(\vec{r})$ and $n(\vec{r})$ treated as independent functional
variables.  If this free energy functional is minimized with respect 
to $n(\vec{r})$ at constant $v(\vec{r})$ (and given $T$, etc.),
the relation
\begin{equation} \label{Euler}
{\frac{\delta F_{\text{HK}}}{\delta n(\vec{r})}}  \; = \;  
- v(\vec{r})
\end{equation}
is obtained.  For $n(\vec{r})$ and $v(\vec{r})$ obeying this physical 
relation, the free energy functional is equal to the grand potential 
by inspection.  The existence of a functional of $n(\vec{r})$ with
this property is one of the basic tenets of DFT, and is the 
(second) Hohenberg--Kohn theorem \cite{HK}.  

Note that below we will 
use \Eq{Euler}, which also follows directly from the properties of
Legendre transforms.  Discussion of the Hohenberg--Kohn 
theorem, \Eq{FEFl}, is nevertheless important even in a 
Legendre--transform--based introduction to DFT, because the 
free--energy--minimization procedure embodied in it is central 
both to forming a physical intuitive picture of DFT, and to 
devising efficient numerical schemes for solving the DFT equations 
in practice \cite{PayneRMP}.

\subsection{Examples}

We next apply the expressions above to two (simplistic) physics 
problems: finding the density distribution of air and of water 
in the ecosphere.

1) \underline{Air:} First we consider air, using the approximate model 
of an ideal gas.  For such a gas the particles do not interact, 
$u(r)=0$, and the partition function can be evaluated directly.  
The Hamiltonian, \Eq{Hamil}, reduces to, 
${\cal H}_{\text MB} = {\cal T} + {\cal V}$.  The grand partition 
function, $\Xi$ of \Eq{Xi_def}, may be expressed as 
$\Xi = \sum_{M=0}^\infty {\Xi_1}^M/M! = \exp(\Xi_1)$, in terms of 
the partition sum for a single particle,
$\Xi_1 = (2\pi\hbar)^{-3} \int {\mathrm d}\vec{r} {\mathrm d} \vec{p} 
\, \exp\LP -(p^2 / 2m +v)/T \RP$.
The momentum integral is trivial, giving $\Xi_1 = 
\int ({\mathrm d}\vec{r} / \lambda_T^3) \exp\LP -v(\vec{r})/T \RP$, 
where volume is normalized by the ``thermal wavelength'' 
$\lambda_T = \sqrt{\hbar^2/mT}$.  One finds from \Eq{GP_def} that
$\Omega = -T \Xi_1$.  In this case the density distribution is 
easily found directly: $n(\vec{r}) = 
\delta \Omega / \delta v(\vec{r}) = \lambda_T^{-3} 
\exp\LP -v(\vec{r})/T \RP$, but for pedagogical purposes we 
proceed to evaluate the Legendre transforms explicitly.

Inverting the ideal gas $n[v]$ relationship just derived gives 
$v(\vec{r}) = -T \log(n \lambda_T^3)$, and proceeding with the 
functional Legendre transform of \Eq{F[n]_def} gives
\beq \label{ig_F}
F_{\text HK}[n(\vec{r})] = \int {\mathrm d}\vec{r} \; n(\vec{r}) \, 
f \LP n(\vec{r}),T \RP  \; ,
\eeq
where
\beq \label{ig_f}
f(n,T) = T \LP\log(n \lambda_T^3) - 1 \RP  \; ,
\eeq
is the free energy per particle.  The latter is equal to the 
Helmholtz free energy per particle, $F(N,T,V)/N$, evaluated for a 
uniform ideal gas with $n=N/V$.

The DFT free energy functional, \Eq{FEFl}, for an ideal gas is 
\beq \label{ig_fefl}
\Omega_v[n(\vec{r})] = \int {\mathrm d}\vec{r} \; 
n(\vec{r}) f\LP n(\vec{r}),T \RP +  
\int {\mathrm d}\vec{r} \; n(\vec{r}) v(\vec{r})  \; .
\eeq
Minimizing this energy functional with respect to $n(\vec{r})$ 
gives $v = -\partial (fn) / \partial n = -T \log(n\lambda_T^3)$,
which is equivalent to $n \propto \exp(-v/T)$, as discussed above.

The density distribution of the atmosphere in the gravitational 
field of the Earth may now be determined.  The external
potential, measured from the chemical potential,
is $v(\vec{r}) = mgh - \mu$, where $h$ is the height and 
$g \simeq 10 {\text m}/{\text s}^2$ is the acceleration due to 
Earth's gravitational field (as we take $h=0$ at Earth's surface,
we have subtracted $\mu$ explicitly here).  Substituting this in 
the expression for the density, we find that the density decreases 
exponentially with height, $n(\vec{r}) \propto \exp(-h/l)$, 
with a length--scale $l = T/mg$.
Taking $m$ as the mass of a nitrogen molecule, 
$m \simeq 5 \times 10^{-26}$kg, and the temperature as 
$T = 4 \times 10^{-21}$ Joules ($\simeq 17^\circ$C), 
gives $l = 8 \times 10^3$m or 8 kilometers, less 
than the height of the Everest (in practice, the atmosphere is 
not in equilibrium, $T \neq {\mathrm const.}$, and this 
description becomes increasingly inaccurate at higher altitudes).

2) \underline{Water:} Finding the density distribution of water 
requires a different analysis, because the ideal gas is a good 
description of real fluids only for low densities such as those 
found in the atmosphere.  At higher densities, the interactions, 
$u(r)$, must be taken into account.  An exact, explicit evaluation 
of the partition function and its derivatives is no longer possible, 
and we are compelled to use models and approximations.  The 
inhomogeneity of the distribution will be given an approximate 
treatment here, 
by assuming that the Hohenberg--Kohn energy may still be expressed 
as in \Eq{ig_F}, in terms of the Helmholtz free energy per particle 
for a uniform system.  This is a ``local density approximation'', 
a concept which is in common use in DFT.  For simplicity, we will 
use the van der Waals model for $f(n,T)$.  In this 
model \cite{vdWaals}, both 
attractive and repulsive interactions of real atoms and molecules
{\bf (}typically included in $u(r)$ for large and small $r$, 
respectively{\bf )} 
are taken into account.  Correspondingly, two simple modifications 
in the ideal--gas expressions for $f$, \Eq{ig_f} are made: first, a 
``higher order in $n$'' attractive term is added, and
second, an ``excluded volume'' $b$, representing the ``hard core''
of real particles, is subtracted from the volume per particle 
$n^{-1}$ in the thermal term.  The Helmholtz free--energy per 
particle in the uniform fluid is thus taken to be 
\beq \label{vdW_f}
f(n)= T \LP 3\log(\lambda_T)-\log(n^{-1}-b) - 1 \RP - an  \; ,
\eeq
which replaces \Eq{ig_f}.  Correspondingly, the pressure is 
\beq \label{vdW_p}
P = -\LP {\partial F \over \partial V} \RP_N = 
- {\partial f \over \partial n^{-1}} = 
{T \over n^{-1}-b} - an^2  \; ,
\eeq
which is a more familiar expression.

The free--energy functional to be minimized with respect to 
$n(\vec{r})$ is again that of \Eq{ig_fefl}, but the extremum condition,
\beq \label{vdWv}
v = -{\partial(n f) \over \partial n} = 
-T \LP {b \over n^{-1}-b} + \log{\lambda_T^3 \over n^{-1}-b} \RP + 2an
\eeq
may now have several solutions, and the value of $n$ which gives 
the lowest free energy must be selected.  In the present 
model, for temperatures which are not too high 
and for a limited range of values of $v$, there are two 
competing local minima, corresponding to the liquid and the gas 
phases of the fluid \cite{instab}.

We now return to water, representing a typical location on Earth 
as a unit of area (one square meter), covered by $10^8$ moles of 
water (one mole $\simeq 6 \times 10^{23}$ molecules).  
We use the same potential $v = mgh-\mu$ and the same 
values of $T$ and $g$ as above.  The mass of a water molecule is 
$m \simeq 18 \times 10^{-3}$kg/mole, and we choose the values of
$a \simeq 0.48$Pa$\times$m$^6/$mole$^2$ and
$b \simeq 16 \times 10^{-6}$m$^3/$mole to 
give the boiling point of water as 100 degrees Celsius at a 
pressure of one atmosphere $\simeq 10^5$Pa, and the particle 
density of the liquid water at that point as 
$n \simeq 56$moles/m$^3$, corresponding to a mass density of 
$10^3$kg/m$^3$.  
Finding the density distribution, $n(h)$, requires the inversion 
of \Eq{vdWv} which is not readily available analytically; therefore 
we have performed this exercise numerically.  The chemical potential 
$\mu$ plays the role of a Lagrange multiplier, imposing the constraint  
$\int {\mathrm d}\vec{r} \, n(\vec{r}) = N$, and finding it 
requires a few trials or iterations (using the Newton--Raphson 
algorithm).

The resulting value of $\mu$ is $-2.5385 \times 10^5$ Joules/mole 
 --- several significant digits must be kept because $v$ (or $v-\mu$) 
varies by only 180 Joules/mole for each kilometer of altitude. 
The density distribution $n(\vec{r})$ contains both liquid and 
gaseous regions.  The liquid region or ``ocean'' occupies the lower 
altitudes, and a gaseous region occupies higher altitudes.  
In the lower region, the fluid is more or less incompressible, with 
the density at $h=0$ higher by only a couple of percent than that 
at ``sea level'', $h\simeq 1750$m.  Above this point one finds 
water vapor with a density of $\simeq 4$moles/m$^3$, corresponding 
to a pressure of $10^4$Pa or one--tenth of an atmosphere (real 
water deviates significantly from our van der Waals model --- a more 
accurate description of $f(n)$ would give $\sim 2.5 \times 10^3$Pa).  
At this low density, the equation of state of the water vapor does 
not differ significantly from that of an ideal gas, and indeed one 
finds an exponential decay with further increase in altitude, as 
discussed above for air (the lengthscale in this case is $\simeq14$km, 
as the water molecules are lighter).

Obviously, water and air coexist on the surface of Earth, and 
taking this into account would modify the results of our examples 
(most importantly, the atmosphere begins only at sea level).  
Such situations can be addressed by generalizing the free--energy 
density, $nf$, allowing it to depend on both densities 
$n_{\text{water}}$ and $n_{\text{air}}$.  With a sufficiently 
accurate parameterization of $f$, a good description of many 
nonuniform thermodynamic systems can thus be obtained.  However, for 
such a local description to hold, the spatial variations in 
$n(\vec{r})$ {\bf (}or equivalently, in $v(\vec{r})${\bf )} must be 
very slow on the scale of the range of the potential $u(\vec{r})$.
In contrast, one may be interested in studying how the density 
distribution $n(\vec{r})$ near a liquid--vapor interface changes 
gradually, on a scale of Angstroms, from that of the liquid to that 
of the gas.  Such problems are typical applications of the DFT of 
classical systems, and must employ nonlocal model functionals 

A discussion of such nonlocal classical functionals is beyond the
scope of the present paper,
but we would like nevertheless to mention the following two points:
(a) As opposed to gases, classical liquids and solids are 
characterized by strong and complicated correlations between 
particles, even when the interaction potential $u(r)$ is simple.  
This makes the task of finding good $F_{\mathrm HK}[n]$ 
functionals nontrivial.  (b) The general framework of DFT can 
nevertheless  give accurate descriptions of real systems, provided 
that one uses good ``anchoring points''.  For example, the 
hard--sphere liquid has been studied extensively numerically, 
and accurate information for it is available.  A good description 
of the liquid--solid transition in several more general systems may 
be found by perturbing in the difference between the actual 
$u(r)$ and that of the hard sphere system, $u_{\text{hs}}(r)$, 
e.g.\ by taking $F_{\mathrm HK}[n] = F_{\text{hs}} + 
\int {\mathrm d}r \> (u-u_{\text{hs}}) 
\LP \delta F_{\mathrm HK} / \delta u(r) \RP$, where in the last 
term, the functional derivative is equal to 
$\delta \Omega / \delta u(r)$ which is the pair correlation function.
With this as background, we now proceed to treat electronic systems 
in the next section.

\section{Application to electrons}

The electron is much lighter than an atom or molecule, 
and thus has a relatively large thermal wavelength, 
e.g.\ $\lambda_T \simeq \hbar/\sqrt{mT} \simeq 17$\AA\ at room 
temperature, more than an order of magnitude larger than the 
typical inter--electron distance.  Its treatment must be 
quantum--mechanical, with
the Hamiltonian of \Eq{Hamil} replaced by the corresponding 
operator, $\hat H_{\text{MB}} = \hat T + \hat V + \hat U$.
The statistical--mechanics derivation above is affected 
by this, e.g.\ the trace ($\text{Tr}$) in the definition of the 
partition function, \Eq{Xi_def}, is taken over the Hilbert space 
rather than over the classical phase space.  However, the 
thermodynamic considerations and Legendre transforms, 
including Eqs.~(\ref{F[n]_def}), (\ref{FEFl}) and (\ref{Euler}) 
for nonuniform systems, are unchanged.

As electrons are Fermions, they form a degenerate gas at the 
prevailing high densities, 
and their energy in the cases of interest here does not 
deviate considerably from its ground--state value.  Indeed, 
electronic DFT was developed in Refs.~\cite{HK} and \cite{KS} 
as a ground--state theory.
Correspondingly, we will from here on take the zero temperature 
limit, $T \to 0$, in all our equations and expressions (cf.\
Ref.~\cite{nolimits}).
In this limit the partition function is dominated by a single 
quantum--mechanical state with an integer number of electrons, 
$M=N$, and the grand potential is equal to 
the ground--state energy (using our $\mu=0$ convention).
The Hohenberg--Kohn free--energy, $F_{\text{HK}}[n]$, is then, 
by \Eq{F[n]_def}, equal to the total energy minus the potential 
energy, i.e.\ to the internal energy.  

For noninteracting quantum mechanical particles, $u(r)=0$, 
the internal energy is just the kinetic energy, for which we 
introduce the notation $F_{\text{ni}}[n] = 
F_{\text{HK}}[n] \left. \vphantom{A \over B} \right|_{u(r)=0}$.
Whereas in the classical case, the noninteracting or ideal--gas 
system had simple local expressions for the energy functionals, 
here the relationships between the energy, the density distribution, 
and the confinement potential are nontrivial and nonlocal.

Practical implementations of DFT require an explicit construction 
of the Hohenberg--Kohn free--energy functional, $F_{\text{HK}}[n]$.  
It is customary to write $F_{\text{HK}}[n]$ for interacting 
electrons as a sum of the noninteracting kinetic energy, 
$F_{\text{ni}}[n]$, 
and two interaction terms --- the electrostatic energy 
and the exchange--correlation energy \cite{differs}: 
\begin{equation} \label{F_sum}
F_{\text{HK}}[n(\vec{r})]  \;  = \;  F_{\text{ni}}[n(\vec{r})] + 
    E_{\text{es}}[n(\vec{r})] + E_{\text{xc}}[n(\vec{r})]  \; ,
\end{equation}
where the last term, $E_{\text{xc}}[n]$, is defined as the remainder and
thus contains everything that is not included in the first two terms. 
Each of the three terms on the right hand side is in principle a 
functional of the independent variable $n(\vec{r})$.  Only the 
second term --- the electrostatic energy --- is easily expressed 
explicitly:
\begin{equation} \label{E_es}
E_{\text{es}}[n(\vec{r})]  \; = \;  {\frac{e^2}2} \int 
 { d\vec r \, d\vec r' \over | \vec r \! - \! \vec r' |}
  \> n(\vec{r}) \, n(\vec{r}^{\prime})  \; ,
\end{equation}
The first and last terms are much more complicated: knowledge of 
the former implies a full understanding of the 
quantum--mechanical noninteracting problem; the latter 
contains all of the many--body physics, and is in principle  
even more complex.

In the following, we briefly introduce two alternative methods for 
confronting this situation: (i) construction of explicit  
approximate expressions for both $F_{\text{ni}}[n]$ and 
$E_{\text{xc}}[n]$, and (ii) the orbital method developed by Kohn 
and Sham, which uses the noninteracting Schrodinger equation to 
evaluate $F_{\text{ni}}[n]$, with only the smaller term, 
$E_{\text{xc}}[n]$, replaced by explicit approximations to the 
desired complicated functional.

\subsection{Explicit functionals}

One of the simplest models of the electronic structure of
atoms was developed by Thomas \cite{Thomas} and Fermi 
\cite{Fermi} already in the late 1920's.  For the noninteracting 
part of the calculation, they used a local density approximation:
\beq \label{TFlda}
F_{\text{ni}}[n(\vec{r})] \simeq C\int n^{5/3}(\vec{r})\>d\vec{r} \; ,
\eeq 
with $C={\frac 3{10}}(3\pi^2)^{2/3}\simeq 2.87$ in atomic units; 
here $C n^{5/3}(\vec{r})$ is the kinetic energy density of 
a {\it uniform} electron--gas of density $n(\vec{r})$. 
They also approximated the Coulomb interaction energy of 
the electrons by the electrostatic term only, 
which corresponds in the present language to taking 
$E_{\text{xc}}[n] = 0$.  Using these simplifications and the 
spherical symmetry of an atom, analytical progress could be made.

Both approximations made in the Thomas--Fermi model are expected 
to be accurate at very high densities.  The approximation made for 
the noninteracting part is accurate when the density of electrons 
changes slowly in space relative to the Fermi wavelength, or 
equivalently, if the density is sufficiently high that 
$n(\vec{r})^{1/3}$ is (much) larger than the spatial rate of 
change of the density.  The approximate treatment of the 
interactions, which ignores exchange and correlation effects, 
is the leading order of a high--density expansion, in powers of 
$r_{\text{s}}/a_0$ where $r_s = (3/4\pi n)^{-1/3}$ is the 
Wigner--Seitz radius, and $a_0=\hbar^2/me^2 = 0.53$\AA\ is the 
Bohr radius.  The Thomas--Fermi model indeed has 
applications for very dense matter, and is  
useful in the description of certain stars \cite{Spruch}.  
Unfortunately, the results for more down--to--earth systems
are rather poor.  For example, it is known that in the Thomas--Fermi
model no molecules can form --- the dissociated atoms always 
have a lower energy \cite{nomol}. 

The Thomas--Fermi method was extended over the years in two main 
directions.  At first, approximate expressions for the 
exchange--correlation energy were included, but this did not lead 
to significant improvements in the results.  Such improvements
were obtained only when an element of nonlocality was taken into 
account in the noninteracting kinetic energy term, $F_{\text{ni}}[n]$.
This was achieved by including gradient terms {\bf (}e.g.\ terms 
proportional to $|\nabla n|^2$ in the integrand of \Eq{TFlda}, with 
$n$--dependent prefactors{\bf )}, and lead in particular to the 
possibility of modeling chemical bonds.  However, the description 
of electronic structure by the Thomas--Fermi model and its 
extensions remains qualitative to date \cite{March}.

\subsection{The Kohn--Sham equations}

In 1965, Kohn and Sham \cite{KS} made a major step towards 
quantitative modeling of electronic structure, by introducing an 
orbital method by which $F_{\text{ni}}[n]$ can be evaluated exactly.
In other words, in order to evaluate the kinetic energy of $N$ 
noninteracting particles given only their density distribution 
$n(\vec{r})$, they simply found the corresponding potential, called 
$v_{\text{eff}}(\vec{r})$, and used the Schrodinger equation,
\begin{equation} \label{KSeq}
\left( -{\frac{\hbar ^2}{2m}}\nabla^2 + v_{\text{eff}}(\vec{r}) 
\vphantom{\half}\right) \psi_i(\vec{r}) = \epsilon_i \psi_i(\vec{r})  
\; ,
\end{equation}
such that $n(\vec{r}) = \sum_{i=1}^N |\psi_i(\vec{r})|^2$.
The states $\psi_i$ here are ordered so that the
energies $\epsilon_i$ are non--decreasing, and the spin index is
included in $i$.  If $\epsilon_N$ is degenerate with 
$\epsilon_{N+1}$ (and also at finite temperatures \cite{finT}), 
fractional occupations $f_i$ are to be used, 
$n(\vec{r}) = \sum_{i=1}^\infty f_i |\psi_i(\vec{r})|^2$, but if 
only spin--degeneracy is involved, the result for the density is 
not affected.  The kinetic energy is then given by 
$F_{\text{ni}}[n(\vec{r})] = 
\sum_{i=1}^N \langle \psi_i | \hat t_i | \psi_i \rangle = 
\sum_{i=1}^N \epsilon_i - \int {\mathrm d}\vec{r} \, n(\vec{r})
v_{\text{eff}}(\vec{r})$, where $\hat t_i$ is the kinetic energy 
operator for the $i$th electron ($\hat T = \sum_i \hat t_i$).

In practice, it is the external potential of a given system which 
is known, not the density distribution or the effective potential.
One may find the effective potential by taking a functional 
derivative of the three--term expression for 
$F_{\text HK}[n]$, \Eq{F_sum}, and rearranging the terms:
\begin{equation} \label{v_eff}
v_{\text{eff}}(\vec{r}) = v(\vec{r}) - e\varphi(\vec{r}) +
v_{\text{xc}}(\vec{r}) \; ,
\end{equation}
where we have used \Eq{Euler}, $\delta F[n]/\delta n = -v$, for 
both the interacting and the noninteracting system.  
The electrostatic potential is here  
\begin{equation}
\varphi(\vec{r}) \; = \; -e \int d\vec{r}^{\prime } \> 
{\frac{n(\vec{r}^{\prime})} {|\vec{r} \! - \! \vec{r}^{\prime }|}} \; ,
\end{equation}
and the exchange--correlation potential is defined as 
\begin{equation} \label{vxc}
v_{\text{xc}}(\vec{r}) \; = \; 
{\frac{\delta E_{\text{xc}}} {\delta n(\vec{r})}}  \; .
\end{equation}
Given a practical approximation for $E_{\text{xc}}[n]$, one obtains
$v_{\text{xc}}(\vec{r})$, and can thus find $v_{\text{eff}}(\vec{r})$
from $n(\vec r)$ for a given $v(\vec{r})$.

The set of equations described above is called the
Kohn--Sham equations of DFT, and must be solved
self--consistently: $v_{\text{eff}}(\vec{r})$ determines 
$n(\vec{r})$ in \Eq{KSeq}, and is determined by it in \Eq{v_eff}. 
They provide a mechanism for minimizing the functional $E_v[n]$ 
(or $\Omega_v[n]$) of 
\Eq{FEFl}, despite the fact that $F_{\text{ni}}[n]$ is known only 
implicitly.  This method assumes that the external potential
$v(\vec{r})$ and the total number of electrons $N$ are given, and
does not use the chemical potential $\mu$ {\bf (}if a nonzero value 
of $\mu$ were restored, it would simply shift both sides of
\Eq{v_eff} by a constant, and thus drop out of consideration{\bf )}.
Together with any explicit approximation for 
the exchange--correlation term (see below), it provides an 
efficient scheme for finding $n(\vec{r})$ and 
the ground state energy \cite{GSenergy} for a system of $N$ 
interacting particles.  Note that for the interacting system, only 
the energy and its derivatives {\bf (}including $n(\vec{r})${\bf )} 
are accessible --- the complicated many--body wavefunctions 
do not take part in this scheme.

Historically, additional properties \cite{IP} of the Kohn--Sham 
noninteracting system, e.g.\ the band structure for crystals, 
have also provided surprisingly accurate predictions when
compared with experiments.  In fact, the agreement between the
calculated Kohn--Sham Fermi surface and the measured one was so
remarkable for some systems \cite{Nick}, that it motivated analyses 
of soluble (perturbative) models for which the 
difference between the
interacting and noninteracting Fermi surfaces could be calculated
explicitly, and shown not to vanish \cite{Mearns}.  Clearly, the
accuracy of DFT predictions for ground--state energies and
density distributions can be improved by finding better 
practical approximations for $E_{\text{xc}}[n]$, whereas improving
the accuracy of such band--structure calculations may require
``going back to the drawing board'' and devising other, more
appropriate, calculational schemes \cite{GW}.

\subsection{The Local Density Approximation}
\label{lda}

\begin{figure}[t]
\epsfxsize=\hsize 
\centerline{\epsffile{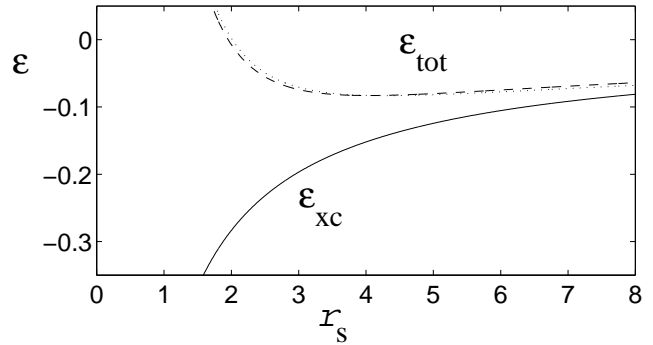}}
\vspace{.3cm}
\caption{The exchange--correlation energy per electron, 
$\epsilon_{\text{xc}}(n)$ (in Hartrees) of a uniform electron 
gas, as a function of the density, parameterized by the 
Wigner--Seitz radius (in Bohr radii), according to the 
interpolation formula of Ref.~{\protect \cite{GL}}.  The dashed 
line is the corresponding total energy per electron, i.e.\ 
includes the kinetic energy.  The Monte--Carlo data, if plotted, 
would be indistinguishable on the scale used here from the curves 
shown.  For reference, we also show the total energy per electron 
(dotted line) according to an alternative parameterisation, the 
well--known Wigner interpolation formula, Ref.~{\protect \cite{Wif}}.
 }
\label{Evsrs}
\end{figure}

As a practical approximate expression for $E_{\text{xc}}[n]$, 
Kohn and Sham \cite{KS} suggested what is known in the context of 
DFT as {\it the} local density approximation, or LDA: 
\begin{equation} \label{tLDA}
E_{\text{xc}}[n(\vec{r})]  \; \simeq \;  \int d\vec{r} \; n(\vec{r})
  \> \epsilon _{\text{xc}}(n(\vec{r}))  \; ,
\end{equation}
where $\epsilon_{\text{xc}}(n)$ is the exchange--correlation energy
per electron in a {\it uniform} electron gas of density $n$.  This
quantity is known exactly in the limit of high density, and can be
computed accurately at densities of interest, using Monte Carlo
techniques (i.e.\ there are no free parameters).  
In practice one usually employs parametric formulas, which are 
fitted to the data and are 
accurate to within 1--2\%.  As an example, we quote
that given by Gunnarson and Lundqvist (Ref.~\cite{GL}): 
$\epsilon_{\text{xc}}(n) = -0.458/r_s - 0.0666 G(r_s/11.4)$
Hartrees, where $r_{\text s}$ is in units of the Bohr radius, 
and $G(x) = {\frac 12}
\{ (1\!+\!x^3)\log(1\!+\!x^{-1})-x^2+{\frac 12}x-{\frac 13} \}$, 
see Fig.~\ref{Evsrs}.

Note that the only difference between the resulting 
computational scheme and a naive mean--field approach is 
the addition of the potential 
\begin{equation}
v_{\text{xc}}(\vec{r}) \; = \; \left.
{d \, \left( n \, \epsilon_{\text{xc}}(n) \vphantom{\half}\right) 
\over dn}\right|_{n = n(\vec r)}
\end{equation}
to the electrostatic potential at the appropriate step in the
self--consistency loop \cite{SIC}.  The corresponding expression for
the ground--state energy is:
\beq
E_0 = \sum_{i=1}^N \epsilon_i  - E_{\text{es}}[n(\vec r)]
     + \int d\vec r \; n(\vec r) \,
  \left( \epsilon_{\text{xc}}(n(\vec r)) - v_{\text{xc}}(n(\vec r))  
    \vphantom{\half} \right)  \; ,
\eeq
where the first term is the noninteracting energy, the second 
term subtracts half of the double counting of the 
electrostatic energy as in the Hartree scheme, and the last term 
is a similar subtraction for the exchange--correlation energy.

\begin{figure}[t]
\epsfxsize=\hsize 
\centerline{\epsffile{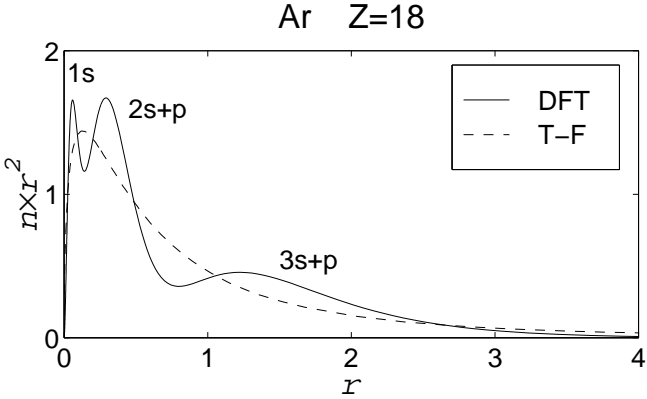}}
\vspace{.3cm}
\caption{ The density $n(r)$ for an Argon atom, multiplied by $r^2$ 
for convenience of presentation (inverse Bohr radii), as a function
of the distance from the origin (in Bohr radii).  The full line is 
the DFT result using the LDA, and shows the shell structure; the 
dashed line is the result of the Thomas--Fermi model for this atom. }
\label{shellstr}
\end{figure}

The LDA has been shown to give very good results for many atomic, 
molecular and crystalline interacting electron systems, even 
though in these systems the density of electrons is not slowly 
varying.  As an example, we show in Fig.~\ref{shellstr} 
the solution for an atom of Argon.  One can see that the shell 
structure, which is absent in the Thomas--Fermi model, is 
described here in detail.  The calculated ground--state energy is 
$-525.9$ Hartrees, compared to $-652.7$ 
Hartrees in the Thomas--Fermi model (one Hartree $= m e^4/\hbar^2 = 
27.2$eV).  The experimental value is $-527.6$ Hartrees --- 
the LDA result is accurate to within less than half a percent, 
compared to the 20\% inaccuracy of the Thomas--Fermi model.  
Both the description of the shell structure and much of the 
improvement in the energy estimate are due to the introduction by 
Kohn and Sham of an exact method for evaluating the highly 
nonlocal kinetic energy functional.
However, the accuracy of the exchange--correlation term is also of 
central importance, and will be discussed next.

\section{Accuracy of the exchange--correlation energy}

In this section, we discuss a few aspects of $E_{\text{xc}}[n]$ which
are somewhat more advanced (and may be skipped in a first reading).
In the first subsection, a formally exact expression for the
exchange--correlation energy is derived, and a sum--rule which
applies to it is obtained.  The next subsection provides references 
to some more recent and accurate approximations, which were devised 
with this sum--rule in mind.
Finally, a general discussion of the level of accuracy achieved by
these approximations for different applications is given.

\subsection{The exchange--correlation hole}

A deeper understanding of the exchange--correlation energy can 
be achieved by considering a continuous transition between the
interacting and noninteracting systems which appear in
the definition of $E_{\text{xc}}[n]$, rather than using a simple
subtraction as in \Eq{F_sum}.  To do this, we reduce the strength 
of the Coulomb interaction, $e^2$, and contemplate a more general 
interaction potential, $u(r)=\Lambda e^2/r$ with 
$0 \leq \Lambda \leq 1$.  In other words, we define the 
Hamiltonian as 
${\hat H}_{\text MB} = {\hat T} + {\hat V} + \Lambda{\hat U}$, 
with the noninteracting system corresponding to $\Lambda=0$, and 
the interacting--electron system to $\Lambda=1$.  For a
given density distribution $n(\vec{r})$ one can consider
$F_{\text{HK}}[n]$ for intermediate values of $\Lambda$,
which leads to the exact expression \cite{Harris}:
\begin{equation} \label{lam_int}
F_{\text{HK}}[n]  \;  = \;  F_{\text{ni}}[n] + 
\int_0^1 {\frac{\partial F_{\text{HK}}[n]}{\partial \Lambda}} 
\> d\Lambda  \; .
\end{equation}
The general rules for Legendre transforms give the derivative 
$\partial F_{\text{HK}}[n] / \partial \Lambda$ as equal to 
${\partial \Omega / \partial \Lambda}$, which in turn is equal to
$\aver{\hat U}$, the expectation value of the interaction energy
${\hat U}$.  Note that the derivative of $\Omega$ is taken at a
constant potential, but in order to reproduce the given
density distribution $n(\vec{r})$, this potential must depend on
$\Lambda$ {\bf (}it is usually denoted by
$v_\Lambda(\vec{r})$, but we need it here only at the two 
extremes, for which we already have appropriate notations:
$v_1(\vec{r}) = v(\vec{r})$ and $v_0(\vec{r}) = 
v_{\text{eff}}(\vec{r})${\bf )}.  

Comparing \Eq{lam_int} with \Eq{F_sum}, we find that
$E_{\text{xc}}[n] = \int_0^1 {\mathrm d}\Lambda \, \aver{\hat U} -
E_{\text{es}}[n]$, i.e.\ the exchange--correlation energy is the
difference between the $\Lambda$--averaged expectation value of the
interaction energy and the electrostatic approximation to it.
Both of these are integrals over the product of the
Coulomb interaction and the density of pairs, which is 
$\half \aver{\hat \rho(\vec r) \LP \hat \rho(\vec r') - 
     \delta(\vec r-\vec r') \RP}$ and $\half n(\vec{r}) n(\vec{r}')$
for the exact and approximate expressions, respectively
(the factor of $\half$ corrects for the double counting of each pair,
and the $\delta$ function removes the interaction of each
electron with itself).  We are thus motivated to define a pair 
correlation function, $g(\vec{r},\vec{r}')$, equal to the 
difference between the two pair densities
\beq \label{rho_xc}
g(\vec{r},\vec{r}')  \; = \;
   \aver{\hat \rho(\vec r) \LP \hat \rho(\vec r') - 
     \delta(\vec r-\vec r') \RP }_{n(\vec{r}),\Lambda}
        - \> n(\vec{r}) \, n(\vec{r}^{\prime})  \; .  
\eeq
It is of relevance here that $g(\vec{r},\vec{r}')$ at 
typical electron densities is relatively featureless --- 
it is only at very low densities ($r_{\text{s}} \sim 100 a_0$) 
that electrons tend to develop strong correlations, and may form 
a Wigner crystal. 

Using the fact that the number of particles, 
$M = \int {\mathrm d}\vec{r}' \, \hat \rho(\vec{r}')$, 
does not fluctuate at very small temperatures, and is equal 
to its expectation value, 
$N = \int {\mathrm d}\vec{r}' \, n(\vec{r}')$, it is easy to 
show that $\int {\mathrm d}\vec{r}' \> g(\vec{r},\vec{r}') = 
-n(\vec{r})$.  It is convenient to define a normalized 
version of this correlation function
$\rho_{\text{xc}}(\vec r, \vec r' ;[n],\Lambda) = 
g(\vec{r},\vec{r}') / n(\vec{r})$, which is the density 
of the so--called exchange--correlation hole.  $\rho_{\text{xc}}$ 
describes the region in $\vec{r}^{\prime}$--space from which 
an electron is ``missing'' if it is known to be at the point 
$\vec{r}$, and the fact that its integral over $\vec{r}'$ is 
equal to $-1$ corresponds to the fact that there is exactly one 
``missing'' electron.  In terms of the $\Lambda$--integrated 
value of this quantity,
\beq
w(\vec{r},\vec{r}') = \int_0^1 {\mathrm d}\Lambda \; 
\rho_{\text{xc}}(\vec r, \vec r' ;[n],\Lambda)  = 
\int_0^1 {\mathrm d}\Lambda \; g(\vec{r},\vec{r}') / n(\vec{r}) \; ,
\eeq
the exchange--correlation energy is given formally as 
\beq \label{F_xc}
E_{\text{xc}}[n(\vec{r})]  \; = \;  
{\frac{e^2}2} \int {\frac{d\vec{r}\,d\vec{r}^{\prime }}
{|\vec{r}\!-\!\vec{r}^{\prime }|}}  \> n(\vec r) \>
w(\vec r, \vec r')  \; .
\eeq
In other words, the exact exchange--correlation energy may be written 
as in the LDA, \Eq{tLDA}, provided that the corresponding energy 
per particle, $\epsilon_{\text{xc}}$ is interpreted not as a local 
quantity but is evaluated according to the density of the
exchange--correlation hole, $\epsilon_{\text{xc}}  = (e^2/2)
\int {\mathrm d}\vec{r}' \, w(\vec{r},\vec{r}') /|\vec{r}-\vec{r}'|$.

The normalization property of the exchange--correlation hole, 
$\rho_{\text{xc}}$, carries over to its $\Lambda$--averaged 
counterpart:
\beq \label{sumrule}
\int {\mathrm d}\vec{r}' \; w(\vec{r},\vec{r}')  = -1  \; .
\eeq
This ``sum--rule'', \Eq{sumrule}, has been used \cite{GJL} as the 
basis for an ``explanation'' of the relatively high accuracy achieved
by the LDA: the reference system here (the homogeneous electron gas)
has properties which are also exact for the inhomogeneous system.
More importantly, it restricts and guides the
search for more accurate practical approximations: expressions
which break the sum--rule can not be expected to work well (see below).

\subsection{Refinements of $E_{\text{xc}}$}
\label{gga}

The fact that the LDA achieves a high {\it relative} accuracy, i.e.\ 
can predict the ground--state energy of various systems to within 
less than a percent, does not mean that the {\it absolute} accuracy 
is sufficient.  In particular, typical applications in chemistry 
require that the energy of a molecule be known to within a small 
fraction of an electron--volt.  Indeed, improving upon the accuracy 
of the LDA is a goal which has been persistently pursued.  One 
improvement which is very often implemented is the
{\it local spin--density} (LSD) approximation \cite{VWN}, which 
is motivated in part by
the fact that the exchange--correlation hole is very different for
electrons with parallel and with antiparallel spins.  In this scheme,
separate densities of spin--up and spin--down electrons are used as a
pair of functional variables: $n_\uparrow(\vec r)$ and
$n_\downarrow(\vec r)$, and the Hamiltonian contains
separate potentials for spin up and spin down electrons --- a
Zeeman--energy magnetic field term is introduced.  The
exchange--correlation energy per particle is then taken from the
results for a homogeneous spin--polarized electron gas,
$\epsilon_{\text{xc}}(n_\uparrow,n_\downarrow)$.  The spin 
dependence allows Hund's rule to be discussed within DFT.

The next degree of sophistication is to allow $\epsilon_{\text{xc}}$
to depend not only on the local densities but also on the
rate--of--change of the densities, i.e.\ to add gradient corrections.
Unfortunately, it was found that such corrections do not necessarily
improve the accuracy obtained.  In fact, introducing gradient 
corrections in a straightforward and systematic manner, by expanding 
around the uniform electron gas, breaks the sum rule of \Eq{sumrule} 
and is less accurate \cite{GJL}.  This situation led to the
development of various {\it generalized gradient approximations}
(GGAs) \cite{GGA,PW91}, in which the spatial variations of
$n(\vec{r})$ enter in a manner which conforms with the sum rule, and
which have succeeded in reducing the errors of the LDA by a factor
which is typically about 4.

Further improvements in practical expressions for $E_{\text{xc}}[n]$
are actively being pursued \cite{conf}.  One direction which may
perhaps achieve the accuracy needed for applications 
in chemistry \cite{Becke}, is to use the fact that the exact form of
the exchange--correlation hole can be calculated for $\Lambda = 0$
relatively easily, directly from the noninteracting Kohn--Sham
system.  There is thus no need to use an approximation such as the LDA
or the GGA for the low--$\Lambda$ portion of the integral in
\Eq{F_xc}.  Ultimately, one hopes that a systematic method of
improving the approximation would be found, although so far this 
has been an elusive goal.

\subsection{Successes and failures}

Over the years, many different types of applications of DFT have 
been developed.  This variety evolved because knowledge of the
electronic ground--state energy as a function of the position of 
the atomic nuclei determines molecular and crystal structure, and 
gives the forces acting on the atomic nuclei when they are not at 
their equilibrium positions.  At present, DFT is
being used routinely to solve problems in atomic and molecular
physics, such as the calculation of ionization potentials
\cite{Gunnarson} and vibration spectra, the study of chemical 
reactions, the structure of bio--molecules \cite{Payne}, and the 
nature of active sites in catalysts \cite{Rajiv}, as well as problems 
in condensed matter physics, such as lattice structures \cite{Heine},
phase transitions in solids \cite{Iron}, and liquid metals
\cite{Gillan}.  Furthermore these methods have made possible the
development of accurate molecular dynamics schemes in which the forces
are evaluated quantum mechanically ``on the fly'' \cite{CarP}.

It is important to stress that all practical applications of DFT
rest on essentially uncontrolled approximations, such
as the LDA discussed above.  Thus the validity of the method is 
in practice established by its ability to reproduce experimental 
results.  A discussion of the accuracy achieved by DFT, compared 
to other alternative approaches, necessarily depends very much on 
the specific applications one has in mind, as detailed below.

For atoms and small molecules, the simplest version of the LDA 
already provides a very useful qualitative and semi--quantitative 
picture.  It is of course a dramatic improvement over the 
Thomas--Fermi model.  It even improves on the more labor--intensive 
Hartree--Fock method in many cases, especially when one is 
calculating the strength of molecular bonds, which are substantially 
overestimated in Hartree--Fock calculations.  This
can only be considered as a surprising success, keeping in mind that
an isolated atom or molecule is as inhomogeneous an electronic system
as possible, and therefore the last place where one might expect a
local approximation to work.  In other words, electronic correlations
in such systems are in a sense weak, and are on average similar to
those of a uniform electron gas {\bf (}see the discussion of the 
sum--rule, \Eq{sumrule}{\bf )}.  However, the many--body quantum states 
of such relatively small systems can be solved for extremely
accurately using well--known techniques of quantum chemistry,
specifically the configuration interaction (CI) method \cite{McW}.
Furthermore, these techniques use {\it controlled} approximations, so
that the accuracy can be improved indefinitely, given a powerful
enough computer, and indeed impressive agreement with experiment is
routinely achieved.  For this reason, most quantum chemists did not
embrace DFT at an early stage.

It is in studies of larger molecules that DFT becomes an indispensable
tool \cite{KBP}.  The computational effort required in the
conventional quantum chemistry approaches grows exponentially with the
number of electrons involved, whereas in DFT it grows roughly as the
third power of this number.  In practice, this means that
DFT can be applied to molecules with hundreds of atoms, whereas 
using CI, one is limited to systems with only a few atoms.  
Simply solving the noninteracting problem for a complicated 
molecule may also be prohibitive, and various methods are used 
in order to reduce the problem to a computationally manageable task.  
Of these, we mention the well--known pseudopotential method
\cite{PayneRMP}, which allows one to avoid recalculating the
wavefunctions of the inert core electrons over and over again, and the
recent attempts to develop ``order N'' methods \cite{ns}, which
make use of the fact that the behavior of the densities at each
point is determined primarily by the atoms in its immediate vicinity,
rather than by the whole molecule. 
It is for this problem that more and more accurate density functionals
are most obviously needed.  To illustrate this, we quote one sentence
from Ref.~\cite{PW91}: ``Accurate atomization energies are found
[using the GGA] for seven hydrocarbon molecules, with a rms error per
bond of 0.1 eV, compared with 0.7 eV for the LSD approximation and 2.4
eV for the Hartree--Fock approximation.''

The remarkable usefulness of DFT for solid--state physics was apparent
from the outset.  For example, the lattice constants of simple
crystals are obtained with an accuracy of about 1\% already in the
LDA \cite{Moruzzi}.  In such applications, the electronic structure of 
a single unit cell with periodic boundary conditions is studied; 
more ambitious applications are also common, e.g.\ a supercell 
containing many unit cells with a single impurity or defect \cite{Makov}.
Admittedly, this method is inappropriate for treating some more
complicated situations, such as antiferromagnets or systems with
strong electronic correlations.  In other cases, such as for the
work--function of metals, local approximations such as the LDA
obviously miss an important part of the physics: for a
point $\vec r$ a short distance away from the surface of a metal, 
the exchange--correlation hole 
$\rho_{\text{xc}}(\vec r, \vec r')$ is concentrated at points 
$\vec r'$ inside or very near the surface of the metal; this
results in image forces, i.e.\ a $1/r$ behavior of $v_{\text{xc}}$ 
(where $r$ is the distance from the surface) which is nonlocal.
However, this deficiency can be corrected for ``by hand'', yielding
satisfactory results \cite{ZLP}.

In general it is useful to note that, in contrast to approximations
using free parameters which are empirically optimized to fit a certain
set of data and may thus be used reliably for interpolation, the LDA
and the GGA have proved to exhibit a consistent degree of accuracy or
inaccuracy for a wide variety of problems --- when applied to a new
problem, the results can thus be interpreted with some confidence.
One should of course also be aware of the cases for which these
approximations are known to fail, such as the image forces mentioned
above, and van der Waals forces \cite{Meir}, which are important 
e.g.\ for biological molecules.  Both of these are manifestations of 
the significance of nonlocal correlations --- a nonlocality which is
by definition absent from the LDA and its immediate extensions.
These examples of practical failure, together with the
unattractiveness of uncontrolled approximations, spur research towards
new and more exact exchange--correlation energy functionals.

Our discussion would not be complete without mentioning the existence
of many other uses of density--functional methods, for electronic
systems and for other physical systems.  The former include
time--dependent DFT, which relates interacting and noninteracting
electronic systems moving in time--dependent potentials, 
and relativistic DFT, which uses the Dirac equation rather than the
Schrodinger equation to calculate the Kohn--Sham states (these are
reviewed in Ref.~\cite{Dreizler}).  The latter include applications in
nuclear physics, in which the densities of protons and neutrons and the
resulting energies are studied \cite{nucl}, and in the theory of
liquids, as already discussed in Sec.~II.

\section{Summary}

In describing density functional theory (DFT) and the 
approximations typically implied by its use, it is necessary 
to follow two steps, as was done in considerable detail in 
Secs.~II and III above.  
The first step is to introduce the Hohenberg--Kohn energy,
$F_{\text{HK}}[n(\vec{r})]$.  It is equal to the internal 
energy, i.e.\ the difference between the ground--state energy 
$E_0$ and the potential energy, or the sum of the kinetic energy 
and the interaction energy of the electrons.  
$F_{\text{HK}}[n]$ is a universal functional of the 
density distribution --- it applies to atoms, 
molecules, crystals, and all other electronic systems.   
The existence of $F_{\text{HK}}[n]$ arises from the fact that
each system has not only a unique external potential,
$v(\vec{r})$, as in traditional many--body theory, but also a 
unique density distribution, $n(\vec{r})$.  
Within DFT, the different systems are labeled by their different 
electronic densities, $n(\vec{r})$, and the potential $v(\vec{r})$ 
is considered as secondary to, and dependent on, the primary 
$n(\vec{r})$.  In fact, the potential is given by the functional 
derivative $v(\vec{r}) = -\delta F_{\text{HK}}/ \delta n(\vec{r})$.
This property of $F_{\text{HK}}[n]$ can be deduced 
in two equivalent ways: (a) it is the Euler equation for the 
Hohenberg--Kohn theorem, which is a variational principle 
stating that the total energy, $F_{\text{HK}}[n] + 
\int {\mathrm d}\vec{r} \> n(\vec{r}) v(\vec{r})$,
is minimized for a given $v(\vec{r})$ by the corresponding 
ground--state density $n(\vec{r})$, and has the ground--state 
energy $E_0$ as its minimum value; 
(b) it arises as the conjugate of the well--known 
relationship $n(\vec{r}) = \delta E_0/\delta v(\vec{r})$, when 
the definition $F_{\text{HK}}[n] = E_0 - 
\int {\mathrm d}\vec{r} \> n(\vec{r}) v(\vec{r})$ 
is viewed as a (functional) Legendre transform.

The functional $F_{\text{HK}}[n(\vec{r})]$ discussed above 
is not known explicitly in terms of its variable, $n(\vec{r})$.  
The second crucial step is to introduce a practical approximation 
for this energy functional.  It may be written as a sum of three 
terms: (i) the kinetic energy term, which is by definition equal to 
the value $F_{\text{HK}}[n(\vec{r})]$ would have for 
noninteracting electrons; (ii) the Hartree term, which is simply a 
double integral over $n(\vec{r})n(\vec{r}')$; and (iii) the remainder 
or exchange--correlation term, which in principle contains all of 
the complicated interaction physics ignored by the first two terms, 
and must be approximated in practice.  By taking a functional 
derivative, one finds that the external potential 
$v(\vec{r})$ is given by a corresponding sum of three terms: (i) the 
so--called effective potential, $v_{\text{eff}}(\vec{r})$, which 
is the potential which a system of noninteracting electrons must 
have in order to reproduce the $n(\vec{r})$ of the interacting 
system, (ii) the electrostatic potential, and (iii) an 
exchange--correlation potential.  The simplest widely--used 
expression for the exchange--correlation term is the local 
density approximation (LDA), which takes the exchange--correlation 
energy--density at each point in the system to be equal to its 
known value for a {\it uniform} interacting electron gas of the 
same density, and results in a parameter--free approximate 
description of all electronic systems.  

These ideas are implemented by the Kohn--Sham 
set of equations, which consists of a noninteracting Schrodinger 
equation involving the effective potential $v_{\text{eff}}(\vec{r})$, 
the abovementioned relationship between $v_{\text{eff}}(\vec{r})$ 
and the given external potential $v(\vec{r})$, and expressions for 
the density distribution $n(\vec{r})$ and the interacting 
ground--state energy $E_0$ in terms of the properties of the 
noninteracting single--particle solutions and of the approximate 
expression for the exchange--correlation energy which is in use.
Using modern computers, the Kohn--Sham equations can be solved 
even for systems containing dozens of atoms, and the results 
for the ground--state energy are typically accurate to within 
a small fraction of a percent.  In contrast to other (much 
more computationally demanding) methods of calculating the 
ground--state energy, the LDA is an uncontrolled approximation, 
and so there is no straightforward path to desired further 
improvements in the accuracy.  Nevertheless, remarkable progress in 
this direction has been achieved over the years, most notably with 
the introduction of the generalized gradient approximation (GGA).

Whereas the outline just given could apply (with minor modifications) 
to other introductions to DFT, the present 
discussion was based on an analogy with thermodynamics.  
It is well known that in treating situations where the the 
number of particles $N$ is constrained, it is preferable to use 
it as a free variable, rather than the chemical 
potential $\mu$.  Similarly, one may think of the 
Coulomb interaction as imposing strong constraints on the 
density distribution $n(\vec{r})$ required to achieve 
low--energy structures in inhomogeneous electronic systems; 
it is thus preferable to use it instead of the potential 
$v(\vec{r})$ as a free variable. 

Three of the advantages of the present approach, as compared, 
e.g., with introducing
DFT using Levy's constrained--search method \cite{Levy}, are: 
(a) the density distribution $n(\vec{r})$ appears here as a natural 
variable --- it is conjugate to $v(r)$ through a Legendre transform  
--- whereas in the conventional description of DFT the very 
existence of the functional $F_{\text{HK}}[n]$ 
appears to be surprising and requires some digestion; (b) some of 
the mathematical difficulties encountered in the ground--state 
theory are not present in the
theory of finite temperature ensembles; and (c) using the standard
properties of Legendre transforms, one immediately obtains the 
physical expression for the exchange--correlation energy in terms 
of the density of the exchange--correlation hole, \Eq{F_xc}, 
an expression which serves as the basis for a discussion of the
weaknesses and strengths of the approximations employed in practice.
We hope that the availability of this type of introduction will help
increase the awareness and understanding of DFT amongst potential 
users, and especially amongst the general audience of physicists and 
scientists.

\subsection*{Acknowledgments}

The authors wish to express their gratitude to N.W. Ashcroft, W. Kohn,
H. Metiu, J.K. Percus, Y. Rosenfeld, and G. Vignale for helpful 
discussions.
N.A. acknowledges support under grants No.\ NSF PHY94-07194, and No.\
NSF DMR96-30452, and by QUEST, a National Science Foundation Science
and Technology Center, (grant No.\ NSF DMR91--20007).


\begin{references}
\bibitem{pointb} $F(N)$ thus defined is not identical with
that obtained by fixing the particle number --- the equivalence of the
canonical and grand--canonical ensembles is guaranteed only in the
thermodynamic limit.  However, we will later focus on the 
$T \to 0$ limit in which the equivalence is regained because the 
fluctuations of the particle number vanish.
\bibitem{instab} Clearly, multiple local minima occur only if 
$-v = \partial(nf)/\partial n = f+nf'$ varies nonmonotonically 
with $n$, i.e.\ possesses regions of negative slope.  This condition
is identical with the conventional one of negative compressibility,
$\kappa<0$, because $\kappa = \partial P/\partial n = 
(\partial/\partial n)(n^2 \, \partial f/\partial n) = 2nf'+n^2f''$, 
which is just $n$ times the abovementioned slope (the density 
$n$ is always positive).
Note that these negative slopes exist only in the phenomenological
model, \Eq{vdW_f}; a proper statistical mechanical evaluation of the
partition function would exhibit regions of phase separation instead.
\bibitem{differs} Note that this definition differs from the
conventional one for exchange--correlation energy, because 
$F_{\text{ni}}[n]$ is the kinetic energy of a system of noninteracting 
particles with the density $n(\vec{r})$, rather than the kinetic 
energy of the interacting electron system.
\bibitem{GSenergy}  For completeness, we write the interacting
ground--state energy $E_0$ explicitly in terms of the Kohn--Sham
eigenvalues and the density distribution:
$$
E_0 = \sum_{i=1}^N \epsilon_i - \int d\vec r \; n(\vec r) \,
  \left( v_{\text{eff}}(\vec r) - v(\vec r) \vphantom{\half} \right) + 
    E_{\text{es}}[n(\vec r)] + E_{\text{xc}}[n(\vec{r})]  \; .
$$
\bibitem{IP}  Interestingly, the work function of a metal surface 
is equal to that of the Kohn--Sham system.  Using the fact that both 
$\varphi(\vec{r})$ and $v_{\text{xc}}(\vec{r})$ decay as $1/r$ at 
large distances away from the system, and using the usual convention of
taking $v(\vec{r})$ and $v_{\text{eff}}(\vec{r})$ to vanish at infinity,
one finds that \Eq{v_eff} reduces to a statement of the equality of the 
chemical potentials for the interacting and the Kohn--Sham system.
\end{references}
\end{document}